\documentstyle[11pt,paspconf,epsf]{article}

\begin{document}

\title{Interstellar Dust in the WIRE to Planck Era}
\author{P. G. Martin}

\affil{Canadian Institute for Theoretical Astrophysics, University of
Toronto, Toronto, Ontario M5S~3H8, Canada}


\begin{abstract}
Interstellar dust appears in a number of roles in the interstellar
medium.  Historically, the most familiar one is as a source of
extinction in the optical.  Absorbed optical and ultraviolet light heats
the dust, whence infrared (including near-infrared to submillimeter)
emission, the particular theme of this review.  In some distant
galaxies, most of the luminosity of the galaxy is thus converted into
the infrared, though in the Milky Way the situation is not generally as
extreme, except in localized regions of star formation.  I briefly
review the range of physical conditions in which the dust emits in the
interstellar medium and the various sizes of dust probed.  Of interest
astrophysically are observations of both the spectrum and the spatial
distribution of the emission, preferably in combination.

In the past fifteen year probes of dust emission have advanced
significantly: through IRAS, KAO, COBE experiments, IRTS, ISO, and MSX.
Satellite and stratospheric observations of dust emission are
complemented by ground-based studies, such as imaging and polarimetry in
the submillimeter with the JCMT or imaging of reflected light from dust
in the near infrared with 2MASS.  Looking ahead, the next decade
promises to be equally exciting.  I give an overview of some of the
distinctive features of facilities anticipated in the near term (WIRE),
early in the new millennium (SOFIA, SIRTF), and somewhat further out
(IRIS, FIRST, Planck).
\end{abstract}


\section{Introduction}

Interstellar dust is fairly cold, emitting in the infrared and
submillimeter, and so as major observational facilities have become
available at these wavelengths unique data have been gathered and great
progress has been made through their analysis.  I describe the basic
processes involved in emission by dust and large molecules.
Observations of both the emission spectrum and the spatial distribution
of the emission are of interest.  I enumerate what data bases are
available now and then turn to the ``candy shop,'' wherein one finds an
enticing array of powerful new facilities that will become available
within the next decade.  The variety of instrumentation offers
broad-band photometric information, higher resolution spectroscopy for
smaller areas of the sky, and wide-field imaging with increased
resolution.  Thus dust in a range of environments from point-like
protostellar environments, to nebulae and more diffuse emission can be
studied.

\section{Emission by Dust}

\subsection{Basic Process}

\subsubsection{Emission by a single grain}
The emissivity of a single grain of size $a$ and temperature $T$ is
\begin{equation} \label{eqsingle}
I_\nu =  \pi a^2 Q_\nu(a)  B_\nu[T(a)],
\end{equation}
where $Q_\nu$ is the efficiency factor for absorption or emission and
$B_\nu$ is the Planck function.  The combination $\pi a^2 Q_\nu$
measures the opacity while $Q_\nu B_\nu$ determines the spectrum.
Appropriate integrations over the size distribution of interstellar dust
are needed.

For $\lambda = c/\nu >> a$, $Q_\nu \propto a$ and so rewriting
equation \ref{eqsingle}
\begin{equation} \label{eqmass}
I_\nu = {4 \over 3} \pi a^3 \,  {3 \over 4} {Q_\nu \over a} B_\nu[T(a)].
\end{equation}
This is the origin of the statement that infrared or submillimeter
emission is a measure of the total mass of dust.  Of course, to
calibrate this one needs $T$ (usually determined empirically from the
broad-band spectrum) and $Q_\nu/a$ (usually from theory, though the
spectral dependence is constrained by observations).  

\subsubsection{Opacity and visual extinction}
With multiwavelength imaging one can obtain temperature-compensated maps
(even without a perfect calibration) of the (relative) spatial
distribution of column density or opacity.  Schlegel, Finkbeiner, \&
Davis (1998) constructed high resolution all-sky maps of optical
extinction $A_V$ using IRAS data (at 4\arcmin\ resolution) with lower
resolution DIRBE and H~I 21-cm emission data and colours of cluster
ellipticals as steps in the calibration (see also Martin 1994 for simple
limited-field application).

\subsection{Grain Temperature}

Grain temperature is determined by radiative energy balance between
absorption and emission (see also \S~\ref{short}).  In the Milky Way the
energy density of starlight and that of the cosmic 3~K microwave
background are about the same and so the radiation temperature is about
3.6~K.  However, this is directly relevant only for a grain with $Q_\nu
= {\rm constant}$, a grey or black body.

Grains are hotter than this because they absorb starlight
(characteristically optical and ultraviolet) fairly efficiently ($Q_\nu
\sim 1$) whereas they emit inefficiently (characteristically infrared;
$Q_\nu << 1$); the grains warm up, increasing $B_\nu$ to make up for the
low $Q_\nu$.  The equilibrium $T$ is the temperature for which the
emission balances the absorption.

A simple illustration is to assume that for all frequencies $Q_\nu
\propto \nu^n$ and that the interstellar radiation field is a blackbody
of temperature $10^4$~K diluted by $W \sim 10^{-14}$, whence $T = 10^4
W^{1/(4+n)}$.  Inserting $n=0$ gives 3.2~K, as mentioned above.
Inserting $n=1$ gives 16~K, very close to that observed in the diffuse
interstellar medium.  But it should be pointed out that the latter is a
coincidence; for the sizes of grains that are actually found in the
interstellar medium the spectral dependences of emission and absorption
give $n$ closer to 2 and 0, respectively, with offsetting effects (the
relative normalization is then important too).  Obviously, in detail $T$
is a function of grain size $a$ and composition even for the same
radiation field.  Some useful examples are given by Draine \& Lee
(1984).

\subsubsection{Different radiation fields}

Dust grains close to luminous stars are bathed in a higher than average
radiation field, and so are hotter.  For example, at a distance of a few
pc from an O8~V star, the dust temperature is closer to 30~K.  On the
other hand, in the quiescent core of a molecular cloud, dust is shielded
from starlight.  A factor of four less radiation to be absorbed would
drop the temperature by $4^{-1/6} = 0.8 \, (n = 2)$, from say 17 to
14~K.

\subsubsection{Infrared absorption}

Where there is a bright infrared background from warm dust (or scattered
light), foreground cooler dust can be imaged in extinction, given
sufficient opacity.  Because of the spectral dependence of $Q_\nu$, the
opacity is larger at higher frequencies.

\subsection{Spectrum} \label{spect}

A good way to assess where grains emit most of their energy is to look
at a plot of $\nu I_\nu$ (or equivalently $\lambda I_\lambda$).  If
$Q_\nu \propto \nu^2$ in the frequency range of emission, then from
equation \ref{eqsingle}, $\nu I_\nu \propto \nu Q_\nu B_\nu \propto
\nu^3 B_\nu$.  This function peaks at
\begin{equation}
\lambda_p \, T = 0.24 \; {\rm cm~K}.
\end{equation}
For high latitude cirrus clouds with $T= 17.5$~K, $\lambda_p = 140 \;
\mu$m (in other units, $\nu_p = 73$~cm$^{-1}$ or $2 \times 10^{12}$~Hz).
Thus IRAS, with the longest wavelength band at 100~$\mu$m (83 --
120~$\mu$m), still missed a substantial fraction of the energy.

\subsubsection{Shorter wavelengths -- non-equilibrium emission}
\label{short} 

\begin{figure}[ht]
\plotfiddle{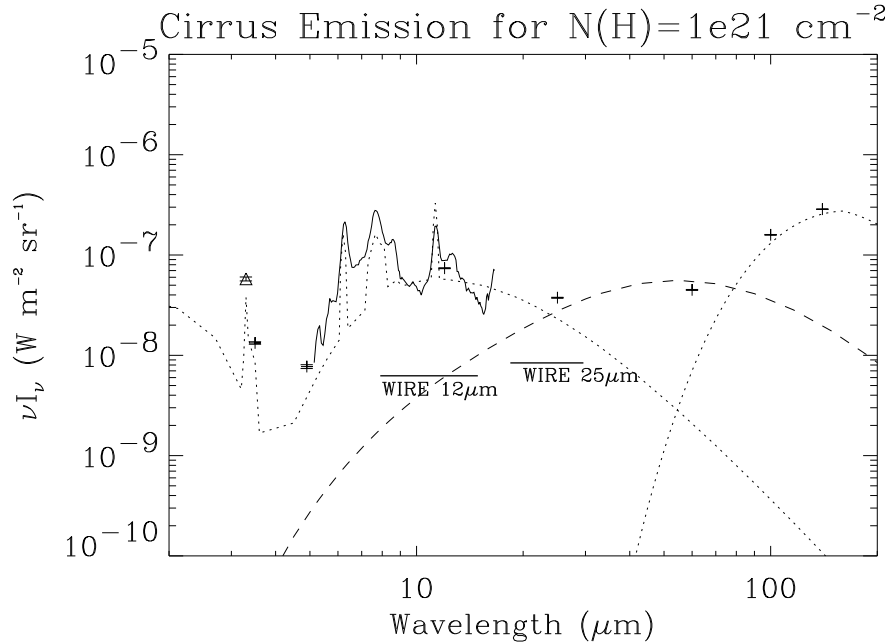}{3.15in}{0}{70}{70}{-208}{-265}
\caption{Emission spectrum of diffuse dust as measured by DIRBE (plus
signs, Bernard et al.\ 1994), the Arome balloon experiment (3.3~$\mu$m
PAH feature, triangle, Giard et al.\ 1994), and ISO (Boulanger et al.\
1996, solid line).  The dotted (near infrared on left), dashed, and
dotted (on right) lines represent emission computed for PAHs, VSGs, and
large grains, respectively, in the model of D\'esert et al.\ (1990). The
horizontal segments for WIRE (see \S~\protect\ref{wire}) mark the 1$ \sigma $
sensitivity at full resolution for a ``minimal overlap'' observing
strategy, which can be improved for faint (lower column density) cirrus
clouds by co-adding.}
\label{fig-1}
\end{figure}

IRAS detected widespread diffuse cirrus emission in the shorter
wavelength bands, at 60, 25, and 12~$\mu$m and this has been followed up
with other experiments (see Fig.~\ref{fig-1}).  Why should there be
emission at $\lambda << \lambda_p$?

The first answer is couched in terms of non-equilibrium emission by very
small grains (VSGs).  These have such a small heat capacity that a
single ultraviolet photon raises the temperature well above the
equilibrium $T$ that would be calculated from the straightforward
radiative balance described above.  To be in this regime, grains must
have $a < 50$~\AA, considerably smaller than classical grains ($a \sim
1000$~\AA) causing optical extinction and polarization, but conceivably
part of a continuous size distribution such as deduced from extinction
(Kim, Martin, \& Hendry 1994).

The grains still emit according to equation \ref{eqsingle}, but now $T$
is a function of time, initially high but decreasing in a matter of
minutes through emission of infrared photons, eventually to below the
equilibrium temperature.  The temperature distribution has been modeled
(e.g., Guhathakurta \& Draine 1989).  VSGs are thought to be responsible
for the enhanced 60 and 25~$\mu$m emission (see model in
Fig.~\ref{fig-1}).

\begin{figure}[ht]
\plotfiddle{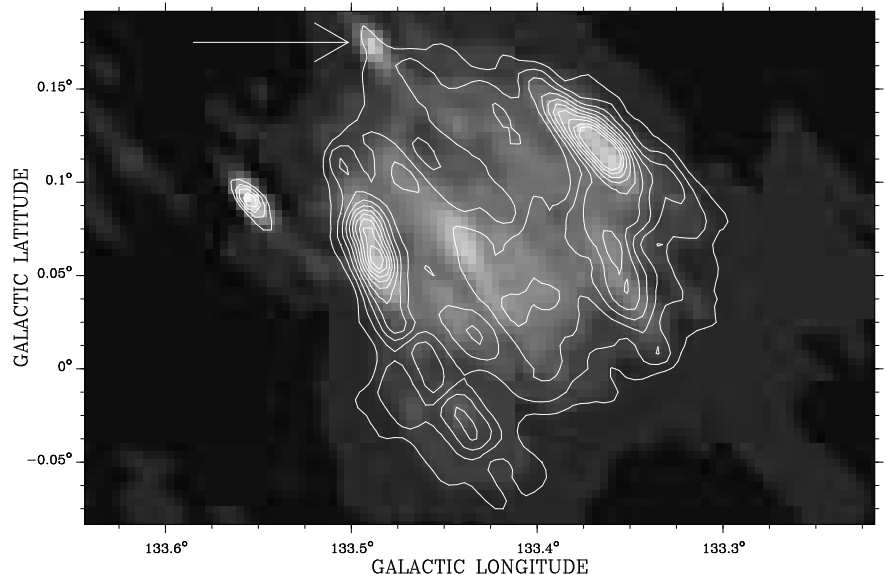}{2.95in}{0}{75}{75}{-220}{-55}
\caption{Spatial distribution of emission from small particles near the
H~II region KR~140.  A 25~$\mu$m IRAS HiRes (enhanced resolution) image
(note the asymmetrical 1\arcmin\ $ \times $ 0.5\arcmin\ beam), showing a
broken shell of emission in the PDR but also revealing VSG emission near
the position of the central exciting star.  Overlaid contours of
12~$\mu$m emission convolved to the 25~$\mu$m resolution show that the
PAH emission extends into the molecular cloud but is absent in the
ionized zone.  The linearly-spaced contour values are 7 -- 25 (in steps
of 2) MJy~sr$^{-1}$.  The arrow points to IRAS 02171+6058 which has
colours of an ultracompact H~II region (the IRAS PSC misidentifies parts
of this bright PDR shell as point sources).  For further discussion, see
Ballantyne, Kerton, \& Martin (1999). } 
\label{fig-2}
\end{figure}

The second answer is more often couched in terms of internal conversion
of the absorbed ultraviolet radiation within large molecules, with
emission in infrared spectral features (if considered as tiny grains,
then the spectral features come from the spectral dependence of
$Q_\nu$).  Prominent in the astronomical setting are polycyclic aromatic
hydrocarbons (PAHs), though the exact species are still poorly
characterized.  Spectral measurements of widespread diffuse cirrus
emission by IRTS and ISO have shown that the 11~$\mu$m PAH feature (see
Fig.~\ref{fig-1}) accounts for a large fraction of the energy detected
in the IRAS 12~$\mu$m band.  Note that PAHs emit a substantial fraction
of $\nu I_\nu$, and so they must account for a corresponding fraction of
the diffuse starlight absorbed.  PAHs are also strong emitters in
photodissociation regions (PDRs) near hot stars where there is an
abundance of ultraviolet radiation (see Fig.~\ref{fig-2}).

\subsubsection{Longer wavelength emission} \label{long}

Emission at $\lambda > \lambda_p$ has been observed with the FIRAS and
DIRBE experiments on COBE.  Early analysis suggested that there was an
excess of longer wavelength emission ($\sim 500$~$\mu$m), attributed to
a relatively cold 7~K dust component (e.g., Dwek et al.\ 1997).  Despite
imaginative suggestions, there is no convincing explanation why some
dust should be this cold.  Recent analysis by Lagache et al.\ (1998) has
concluded that the 7~K dust component is spurious, resulting from
analyses which neglected an isotropic far infrared background.  This
background has a different spectrum and causes significant contamination
at long wavelengths (they also find that the long wavelength Galactic
dust spectrum is consistent with $n = 2$).

At even lower frequencies, in the microwave range (15 -- 50~GHz), there
is another excess over the thermal emission expected from dust, but
which correlates with 100~$\mu$m emission.  Lazarian (this conference;
also Draine \& Lazarian 1998) attributes this to emission from small
spinning dust grains.  This component of emission is another
``foreground,'' possibly polarized, that has to be subtracted in
studying the cosmic microwave background (\S~\ref{planck}).

\section{Web Sites}

There are many sources of infrared and submillimeter data on dust.  Much
more information that what can be given here is available via the
respective web sites, a compilation of which is given below.  To provide
points of reference, I have listed here many sites for ``completed'' to
``on-going'' facilities.  In the next section, I touch on a few points
for ``future'' missions.

\subsection{Largely Complete}

\begin{enumerate}

\item IRAS: http://www.ipac.caltech.edu/ipac/iras/iras.html

\begin{itemize}

\item HIRES: http://www.ipac.caltech.edu/ipac/iras/hires\_over.html

\item IGA: http://irsa.ipac.caltech.edu/applications/IGA/

\item MIGA: http://www.cita.utoronto.ca/$\sim$kerton/ \hfil\break
\null  $\qquad \quad$ (see also Kerton \& Martin, this conference)

\end{itemize}

\item COBE: http://www.gsfc.nasa.gov/astro/cobe/cobe\_home.html

\item IRTS: http://koala.astro.isas.ac.jp/irts/irts\_E.html

\item KAO: http://ccf.arc.nasa.gov/dx/basket/storiesetc/FSARC004.html and \hfil \break 
\null $\qquad \, \, \,$ http://jean-luc.arc.nasa.gov/KAO/homepage.html

\end{enumerate}

\subsection{Significant Data Still to be Released}

\begin{enumerate}

\item ISO: http://www.iso.vilspa.esa.es/ $\quad$ and \hfil \break
\null $\qquad$ http://www.ipac.caltech.edu/iso/iso.html 

\item MSX: http://gibbs1.plh.af.mil/

\end{enumerate}

\subsection{On-going Ground Based}

\begin{enumerate}

\item JCMT: http://www.jach.hawaii.edu/JCMT/

\begin{itemize}

\item SCUBA: http://www.jach.hawaii.edu/JCMT/scuba/

\end{itemize}

\item 2MASS: http://www.ipac.caltech.edu/2mass/

\end{enumerate}

\section{Future}

There are many upcoming missions and facilities of great interest to the
study of dust.  Again, there a web sites to find basic information and
keep current.

\begin{enumerate}

\item WIRE: http://www.ipac.caltech.edu/wire/

\item MAP: http://map.gsfc.nasa.gov/

\item SOFIA: http://sofia.arc.nasa.gov/

\item SIRTF: http://ssc.ipac.caltech.edu/sirtf/

\item IRIS: http://koala.astro.isas.ac.jp/Astro-F/index-e.html

\item PLANCK: http://astro.estec.esa.nl/SA-general/Projects/Planck/

\item FIRST: http://astro.estec.esa.nl/SA-general/Projects/First/

\end{enumerate}

IRIS and also MAP and Planck will survey the whole sky like their
predecessors IRAS and COBE.  The others will observe more limited areas
with some combination of higher sensitivity, spatial resolution, and
spectral resolution.  Issues in evaluating and intercomparing the
upcoming prospects include size of mirror, field of view, array
detectors, imaging and spectroscopic capabilities, sensitivity and noise
(cooling; backgrounds), expected availability, service lifetime, and
access to observations and data.  The various web pages provide some
useful intercomparisons with predecessors and competitors.

\subsection{Wide Field Infrared Explorer -- WIRE} \label{wire}

WIRE, a small 30-cm telescope in NASA's explorer series, is to be
launched in the spring of 1999.  Because the telescope is cooled (with
solid H$_2$), the mission lasts only about four months.  The primary
goal of WIRE is to reveal how galaxies evolve with time at infrared
wavelengths.  During repointing from one Galactic pole to another (about
18\% of clock time) WIRE will make sensitive observations toward regions
that do not compete with the primary mission.  This time has been
awarded to a number of Associate Investigator (AI) programs.  AI data
are proprietary for a year and then will be released in an archive.

WIRE produces images at 12 and 25~$\mu$m (see Fig.~\ref{fig-1}) with
about 20\arcsec\ resolution over a 33\arcmin\ field.  Thus it is
possible to map significant portions (several square degrees) of
interstellar clouds.  A team headed by G. Helou with which I am involved
will study the cloud structure and relative distribution of emission of
the distinct PAH and VSG components, providing important information on
their evolution and sensitivity to local conditions in a variety of
environments (molecular, translucent, cirrus).  Dense cores will be
detected in extinction.  We shall also examine the evolution of dust in
interfaces between H~II regions and molecular clouds (e.g.,
Fig.~\ref{fig-2}).

\subsection{Stratospheric Observatory for Infrared Astronomy --SOFIA}

A follow-on to the Kuiper Airborne Observatory (KAO), SOFIA is a Boeing
747-SP aircraft modified to accommodate a 2.5-m telescope.  Developed by
NASA and DLR, it is expected to begin flying by mid-2001 and have an
operating lifetime of some 20 years with about 160 flights (about 5
hours actual observing each) per year.  Initially it has four facility
class science instruments for the use of the user community (via peer
review) who will not be required to have extensive knowledge or
experience in infrared instrumentation or observing techniques.  HAWC
(high resolution airborne wideband camera) with a $12 \times 32$ array
covers the 40 -- 300~$\mu$m spectral range with three bands centered at
60, 110, and 200~$\mu$m at an image scale of two pixels per Airy disk
(FWHM; diffraction limited resolution about 15\arcsec\ at 110~$\mu$m).
FORCAST (a wide-field infrared camera) will sample at 0.75\arcsec /pixel
over a 3.2\arcmin\ $\times$ 3.2\arcmin\ instantaneous field of view for
diffraction limited imaging.  Selectable filters provide for continuum
imaging in the 16 -- 40~$\mu$m region.  Clearly, the increased spatial
resolution of these instruments is coming at the expense of sky
coverage.  AIRES (airborne infrared echelle spectrometer) utilizes
two-dimensional detector arrays and a large echelle grating to achieve
spectral imaging with high angular resolution and sensitivity.
Wavelength coverage can be selected within 17 -- 210~$\mu$m. FLITECAM
(first light infrared test experiment camera) will offer wide-field
imaging, high-resolution imaging for observing diffraction-limited
images at 3~$\mu$m, low-resolution grism spectroscopy, and
pupil-viewing.

There are also five principal investigator class first light infrared
instruments, which are all spectrometers (those with resolving power $R
= \lambda/\Delta\lambda > 100$ are probably less interesting for studies
of dust).  Given the long lifetime of the observatory, it will be
possible to make instrumentation improvements taking advantage of new
technologies like advanced large format arrays.

\subsection{Space Infrared Telescope Facility -- SIRTF}

SIRTF, the last of the ``Great Observatories'' of NASA, consists of an
85-cm telescope.  Despite a dramatic reduction in budget, the
scientific potential has been maintained.  The mission cryogenic
lifetime is 2.5~y, though 5~y is possible.  Launch is in late 2001 into
a solar earth-trailing orbit.  Like HST there will be GTO time, but a
large portion will be for General Observer peer-reviewed proposals.  In
addition there is Legacy Science, a unique category of peer-reviewed
programs to be carried out by teams of investigators.  Legacy Science
programs are distinguished as being large (perhaps more than a thousand
hours) coherent science investigations that produce a scientific data
archive which is of general and lasting importance to the broad
community (such data are non-proprietary).  These large projects are
expected to comprise a significant fraction of SIRTF's first year of
science operations.

SIRTF has three cryogenically-cooled science instruments for imaging and
spectroscopy over 3 -- 180~$\mu$m, with large sensitivity gains by using
large-format detector arrays.  MIPS (multiband imaging photometer for
SIRTF) is comprised of three detector arrays: $128 \times 128$ for
imaging at 24~$\mu$m, $32 \times 32$ at 70~$\mu$m, and $2 \times 20$ at
160~$\mu$m; the fields of view are 5\arcmin, 5\arcmin, and 0.5\arcmin\ $
\times 5$\arcmin, respectively.  There is a scan mirror to provide
mapping of larger fields with very efficient use of telescope time.  The
$32 \times 32$ array also takes very low-resolution spectra ($R = 10$)
from 50 -- 100~$\mu$m.  IRAC (infrared array camera) has four channels
providing simultaneous images at 3.6, 4.5, 5.8, and 8~$\mu$m; the $128
\times 128$ arrays give a 5\arcmin\ field of view with 1.2\arcsec\
pixels.  IRS (infrared spectrograph) has two modules providing spectra
with $R \sim 50$ over 4 -- 40~$\mu$m and another two providing $R \sim
600$ over 10 -- 37~$\mu$m.  The low-resolution modules are long slit
designs that allow both spectral and one-dimensional spatial information
to be acquired simultaneously on the same detector array.  The
high-resolution modules use a cross-dispersed echelle design that gives
both spectral and limited spatial measurements on the same detector
array.

\subsection{Infrared Imaging Surveyor -- IRIS}

IRIS, a 70-cm cooled telescope, is the second infrared astronomy mission
(after IRTS) of ISAS with a launch in early 2003.  Unlike many of the
other facilities being described, IRIS will carry out a full infrared
sky survey with FIS (far-infrared surveyor).  FIS is sensitive in the
range from 50 -- 200~$\mu$m, with angular resolutions of 30 --
50\arcsec\ and much greater sensitivity than IRAS.  There are four
passbands: 50 -- 70, 50 -- 110, 150 -- 200, 110 -- 200~$\mu$m; the
longer wavelength coverage is important for diffuse dust
(\S~\ref{spect}).  For pointed observations FIS also incorporates a
Fourier spectrometer covering the entire range, with $R = 200$ at
100~$\mu$m.

IRC (infrared camera) has large-format detector arrays to take deep
images of selected regions (field of view 10\arcmin) with spatial
resolution about 2\arcsec; three channels observe simultaneously to
cover the ranges 1.8 -- 5, 5 -- 12, and 10 -- 26~$\mu$m.  There is also
a capability of using grisms for low-resolution ($R \sim 50$) slitless
spectroscopy.

\subsection{Planck/FIRST} \label{planck}

I lump these two very distinct ESA satellites together because they have
been combined for a 2007 ``carrier'' launch and deployment to orbits
around the second earth-sun Lagrange point (L2).  Planck is a 1.5-m
telescope for all-sky mapping of the cosmic microwave background.  Full
sky coverage is obtained in six months.  Interstellar medium studies
come as a by-product of having to subtract the Galactic foregrounds
(free-free and synchrotron continua as well as dust).  Wide frequency
coverage will aid the separation of the various components which have
distinctive spectra.  HFI (high frequency instrument) will have spectral
coverage with bolometers in six channels ($R=4$) from 350~$\mu$m to 3~mm
(860 -- 100~GHz) at 5\arcmin\ resolution above 200~GHz (like the ISSA
product from IRAS).  HFI also measures linear polarization in three
channels.  The polarized dust foreground in this frequency range will be
from thermal emission (\S~\ref{long}).  Polarized emission from rapidly
spinning small grains will be of interest to the LFI, with HEMT
receivers giving spectral coverage over 3 -- 10~mm (100 -- 30~GHz) in
four ($R=5$) channels and spatial resolution as good as 2\arcmin\ at
40~GHz.  Note also MAP (NASA's microwave anisotropy probe; much earlier
launch in fall 2000) which will cover 22 -- 90~GHz with a highest
resolution of 13\arcmin.

FIRST (far infrared and submillimeter telescope), a 3.5-m passively
cooled telescope for pointed studies, has an optimum range 80 --
670~$\mu$m for photometry and spectroscopy, considering complementary
facilities in space (SIRTF) and on the ground (SCUBA).  FIRST is quite
sensitive, one rough comparison being a yearly output equivalent to
about 1000 SOFIA flights (with SOFIA first light instrumentation).
Eventually 3/4 of the time will be available to general observer
proposals.  There are three planned instruments (about to be confirmed),
with cryostat lifetime about 3~y.  Of most interest to studies of dust
are PACS and SPIRE.  PACS (photoconductor array camera and spectrometer)
simultaneously covers 80 -- 130~$\mu$m and 130 -- 210~$\mu$m with $25
\times 16$ arrays providing 1.5\arcmin\ $\times$ 1\arcmin\ and 3\arcmin\
$\times$ 2\arcmin\ coverage, respectively, at full beam sampling.  The
spectroscopic capability provides $R \sim 1800$.  SPIRE (spectral and
photometric imaging receiver) has a 4\arcmin\ field for simultaneous
photometry ($R = 3$) at 250, 350, and 500~$\mu$m using square arrays of
bolometers with 32, 24, and 16 on a side, respectively.  It includes an
FTS with a 2\arcmin\ field of view for spectroscopy (adjustable to $R =
20$ -- 1000 at 250~$\mu$m) over 200 -- 670~$\mu$m.  HIFI is for high
resolution ($R \sim 10^3$ -- $10^6$) heterodyne spectroscopy in three
bands within 480 -- 2700~GHz (625 -- 110~$\mu$m).

\acknowledgments This work was supported by the Natural Sciences and
Engineering Research Council of Canada.

\vfill


\begin{references}
\reference Ballantyne, D. R., Kerton, C. R., \& Martin, P. G. 1999,
\apj, in preparation 
\reference Bernard, J. P., Boulanger, F., D\'esert, F. X., Giard, M., Helou, G.
and Puget J. L. 1994, A\&A, 291, L5
\reference Boulanger, F., et al.\ 1996, \astap, 315, L325
\reference D\'esert, F-X., Boulanger F., Puget J. L. 1990, A\&A, 237, 215
\reference Draine, B. T., \& Lazarian, A. 1998, \apj, 508, 157
\reference Draine, B. T., \& Lee, H. M. 1984, \apj, 285, 89
\reference Dwek, E., et al.\ 1997, \apj, 475, 565
\reference Giard, M., Lamarre, J. M., Pajot, F., Serra, G. 1994, \astap, 286, 203
\reference Guhathakurta, P. \& Draine, B. T.  1989, \apj, 345, 230 
\reference Kim, S.-H., Martin, P. G., \& Hendry, P. D. 1994, \apj, 422, 164
\reference Lagache, G., Abergel, A., Boulanger, F., \& Puget, J.-L. 1998,
\aap, 333, 709
\reference Martin, P. G. 1994, in Infrared Cirrus and Diffuse
Interstellar Clouds, R.\ M. Cutri and W.\ B. Latter, ASP Conference
Series 58: San Francisco, 137
\reference  Schlegel, D. J., Finkbeiner, D. P., \& Davis, M. 1998, \apj,
500, 525 
\end{references}
\end{document}